\documentclass[11pt]{article}
\usepackage{latexsym}
\usepackage{amssymb}
\usepackage{amsmath}
\usepackage{subfigure}
\usepackage[dvips]{graphicx}

\title{Solutions of $D=2$ supersymmetric Yang-Mills quantum mechanics with $SU(N)$ gauge group}
\author{Piotr Korcyl}
\date{M. Smoluchowski Institute of Physics, Jagiellonian University, Reymonta 4, 30-059 Krak\'{o}w, Poland}

\begin{document}

\maketitle

\begin{abstract}
We describe the generalization of the recently derived solutions
of $D=2$ supersymmetric Yang-Mills quantum mechanics with $SU(3)$
gauge group to the generic case of $SU(N)$ gauge group.
We discuss the spectra and eigensolutions in bosonic as well as fermionic sectors.
\end{abstract}


        \section{Introduction}

        Supersymmetric Yang-Mills quantum mechanics \cite{witten,claudson} attract a lot of attention since the works Ref.\cite{banks,hoppe}.
        Among many variants of such quantum mechanics, the twodimensional
        systems are the simplest ones. Although proposed nearly 30 years ago by Claudson and Halpern \cite{claudson}, their solutions for gauge
        groups other than $SU(2)$ are poorly known. Recently, complete solutions were derived for the model with the $SU(3)$
        gauge group \cite{korcyl4}. In this work we present a generalization of such solutions to the case of models with any $SU(N)$ gauge group.
        We discuss their derivation as well as their properties. We use the framework of cut Fock basis described in details in Ref.\cite{korcyl1}, which
        allows a systematic analysis of the SYMQM systems by both numerical and analytic methods.

        The cut Fock space
        approach basically consists in introducing the Fock basis in the Hilbert space and then
        in considering only a finite subset of basis states. This subset is composed of states with less than
        $N_{cut}$ quanta and $N_{cut}$ is usually called the cut-off.

        The main results presented in this paper are
        the closed formulae for the eigenenergies and corresponding eigenstates of
        SYMQM in any, bosonic or fermionic, sector, valid for any $N_{cut}$. Their infinite cut-off limit is also discussed.

        Such results are especially interesting since the enable one to study the large-$N$ behavior of the wavefunctions
        of SYMQM as well as the supersymmetric structure of the limiting model. They may be used to construct
        higher dimensional wave-functions which can have an interpretation in the context of supermembrane theory \cite{maciek_spiky}.
        One may also try to build a perturbation theory around their finite cut-off versions \cite{korcyl1}.

        This paper is composed in the following way. We start by briefly presenting the $D=2$ supersymmetric Yang-Mills quantum mechanics
        and the approach of cut Fock basis. Then, we translate the eigenequation for the Hamiltonian to a recursion relation for the
        coefficients
        describing the decomposition of the eigenstate in the Fock basis. The recursion relation is valid for any $N$.
        Subsequently, we discuss the implications of this recurrence relation. Specifically, we argue that a closed formula for the spectra
        can be deduced in all, bosonic as well as fermionic, sectors of the SYMQM models. We explicitly present it for
        the $SU(4)$ and $SU(5)$ models. We also mention the form of the spectra in the infinite cut-off limit. In section \ref{sec. solutions} expressions for
        the eigenstates are presented. The emphasis is put on the wave-functions of supersymmetric vacua in a generic $SU(N)$ model.
        Next, the completeness, orthogonality, normalization and infinite cut-off limit of our eigensolutions
        are discussed. Finally, we end with some conclusions.

        \section{Supersymmetric Yang-Mills Quantum Mechanics}
        \label{sec. symqm}

        Although supersymmetric Yang-Mills quantum mechanics were described already many times \cite{claudson,wosiek1,doktorat_macka,korcyl1},
        let us briefly remind the main statements in order to keep this work self-contained.
        SYMQM can be obtained by a dimensional reduction of a supersymmetric, $D=d+1$ dimensional Yang-Mills quantum
        field theory to $D=0+1$, i.e. to a single point in space. The remaining degrees of freedom are those of internal symmetries
        of the field theory. Consequently, the initial local gauge symmetry is reduced to a global symmetry of the quantum mechanical system.
        In this work we will be interested in systems obtained from
        $\mathcal{N}=1$ Yang-Mills field theory in two dimensions with different $SU(N)$ gauge symmetries \cite{claudson}. The
        degrees of freedom are described by a scalar field $\phi_A$
        and a complex fermion $\lambda_A$, where $A$ labels the generators of the gauge group, hence $\phi$ and $\lambda$ transform in the
        adjoint representation of $SU(N)$. The Hamiltonian of the reduced system reads \footnote{The summation of doubled indices is understood, i.e. $\phi_A \phi_A \equiv \sum_{A=1}^{N^2-1} \phi_A \phi_A$.}
        \begin{equation}
        H = \frac{1}{2} \pi_A \pi_A + i g f_{ABC} \bar{\lambda}_A \phi_B \lambda_C.
        \end{equation}
        $H$ is supersymmetric since we can define the supercharges $Q$ and $\bar{Q}$,
        \begin{equation}
        Q = \lambda_A \pi_A, \qquad \bar{Q} = \bar{\lambda}_A \pi_A,
        \end{equation}
        such that
        \begin{equation}
        \{ Q, Q \} = \{ \bar{Q}, \bar{Q} \} = 0, \quad \textrm{ and } \quad \{ Q, \bar{Q} \} = \pi_A \pi_A - 2 g \phi_A G_A,
        \end{equation}
        where
        \begin{equation}
        G_A = f_{ABC}\big( \phi_B \pi_C - i \bar{\lambda}_B \lambda_C \big),
        \end{equation}
        is the generator of the gauge transformations.
        Upon canonical quantization, one defines operators satisfying canonical commutation/anticommutation relations,
        \begin{equation}
        [\phi_A, \pi_B] = i \delta_{A,B}, \qquad \{ \lambda_A, \bar{\lambda}_B \} = \delta_{A,B}.
        \end{equation}
        The quantization procedure requires the imposition of the Gauss' constraint on physical states. The dimensionally
        reduced Gauss' law translates to a requirement of invariance of physical states under gauge transformations,
        \begin{equation}
        G_A |\textrm{ physical state } \rangle = 0.
        \end{equation}
        Thus, the fermionic term of the Hamiltonian, being proportional to the Gauss' constraint, vanishes on any physical state.
        Therefore,
        \begin{equation}
        H = \big\{ Q, Q^{\dagger} \big\} = \frac{1}{2} \pi_A \pi_A \qquad \textrm{ in the physical Hilbert space}.
        \label{eq.hamiltonian}
        \end{equation}
        We end this section by rewriting the above Hamiltonian in terms of creation and annihilation operators, defined as
        \begin{equation}
        a_A = \frac{1}{\sqrt{2}}\big( \phi_A + i \pi_A\big), \qquad a^{\dagger}_A = \frac{1}{\sqrt{2}}\big( \phi_A - i \pi_A\big).
        \end{equation}
        Thus,
        \begin{equation}
        H = \textrm{tr} (a^{\dagger}a) + \frac{N^2-1}{4} -\frac{1}{2} \textrm{tr} (a^{\dagger} a^{\dagger}) - \frac{1}{2}\textrm{tr} (aa),
        \label{eq.hamiltonian2}
        \end{equation}
        where we used the matrix notation, in which every operator transforming in the adjoint representation
        is summed with the generators of the $SU(N)$ group in the fundamental representation, giving an operator valued $N \times N$ matrix.
        In what follows we will use a simplified notation for the trace of any such matrix, namely, $\textrm{tr}(O) \equiv (O)$.

        \section{Gauge invariant Fock basis for $SU(N)$ SYMQM}

        A systematic construction of the Fock basis for the $D=2$ SYMQM models
        was proposed in Ref.\cite{wosiek1} and developed in Ref.\cite{doktorat_macka}. We use the notation
        introduced in Ref.\cite{korcyl1} where the recursive construction of the basis using
        the notion of elementary bosonic and fermionic bricks was described in details.
        We recall here very briefly the most important conclusions
        concerning the Fock basis of the SYMQM Hilbert space.

        An operator is called an elementary bosonic brick if it is a single trace operator composed exclusively
        of creation operators. For a given $N$ we have $N-1$ linearly independent elementary bosonic bricks,
        which we label by $C^{\dagger}$. They are
        \begin{equation}
        C^{\dagger}_N(2) \equiv (a^{\dagger 2}), \ C^{\dagger}_N(3) \equiv (a^{\dagger 3}),
        \ \dots, \ C^{\dagger}_N(N-1) \equiv (a^{\dagger N-1}), \ C^{\dagger}_N(N) \equiv (a^{\dagger N}). \nonumber
        \end{equation}
        A generic basis state can be written as
        \begin{equation}
        |p_2, p_3, \dots, p_N\rangle =  C^{\dagger}_N(2)^{p_2} C^{\dagger}_N(3)^{p_3}  \dots  C^{\dagger}_N(N-1)^{p_{N-1}}  C^{\dagger}_N(N)^{p_N} |0\rangle.
        \label{eq. bozonowy stan bazowy sun}
        \end{equation}

        Additionally, in the fermionic sector with $n_F$ fermionic quanta there are $d^{n_F}(N)$ fermionic bricks.
        We label them by $C^{\dagger}_N(n^{\alpha}_B, n_F, \alpha)$, where $n^{\alpha}_B$ denotes the number of bosonic
        creation operators and $n_F$ the number of fermionic creation operators incorporated in $C^{\dagger}_N(n^{\alpha}_B, n_F, \alpha)$.
        $\alpha$ is an additional index, since $n^{\alpha}_B$ and $n_F$ do not specify unambiguously the operator. $\alpha$ runs from $1$
        to $d^{n_F}(N)$ in each fermionic sector. Fermionic basis states can be obtained by the application of the fermionic bricks to the bosonic
        basis states eq.\eqref{eq. bozonowy stan bazowy sun}. Hence, we define
        \begin{equation}
        |\alpha, n_F; p_2, p_3, \dots, p_N\rangle = C^{\dagger}_N(n^{\alpha}_B, n_F, \alpha)|p_2,p_3,\dots,p_N\rangle.
        \label{eq. fermionowy stan bazowy sun}
        \end{equation}
        Although the sets of fermionic bricks are not explicitly known for $N>4$, it
        turns out that they are not necessary for the derivation of spectra of the $SU(N)$ SYMQM models.

        A generic state from the bosonic sector with up to $N_{cut}$ bosonic quanta can be decomposed as
        \begin{equation}
        |E\rangle = \sum_{2 p_2+ 3 p_3+ \dots + N p_N \le N_{cut}} a_{p_2, p_3, \dots, p_N} |p_2, p_3, \dots, p_N \rangle,
        \label{eq. dekompozycja stanu bozonowego}
        \end{equation}
        whereas a generic state from the sector with $n_F$ fermionic quanta can be
        decomposed in the so constructed basis with unknown amplitudes
        $a^{\alpha}_{p_2, p_3, \dots, p_N}(E)$, where
        the index $\alpha$ describes which one of the fermionic bricks was used. For the cut-off $N_{cut}$ we get
        \begin{equation}
        |E\rangle = \sum_{\alpha=1}^{d^{n_F}(N)} \sum_{\sum_{k=2}^N k p_k \le N_{cut}-n^{\alpha}_B} a^{\alpha}_{p_2, p_3, \dots, p_N} |\alpha,n_F;p_2, p_3,\dots, p_N \rangle.
        \label{eq. dekompozycja stanu fermionowego}
        \end{equation}

        The physical results correspond to the limit of $N_{cut} \rightarrow \infty$. Such limit is nontrivial in the case of systems
        with continuous spectra and was discussed in details in the case of a system with one degree of freedom \cite{maciek1}
        as well as systems with a $SO(d)$ gauge symmetry \cite{korcyl0}. In this paper we will describe the $N_{cut} \rightarrow \infty$
        limit, however, the study of the scaling law in the spirit of Refs.\cite{maciek1,korcyl0} will be discussed elsewhere.


        \section{Recurrence relations}

        In this section we derive the recurrence relation for the coefficients $a_{p_2, p_3, \dots, p_N}$ and $a^{\alpha}_{p_2, p_3, \dots, p_N}$.
        To this goal we follow the derivation of a similar recurrence relations for the $SU(3)$ model \cite{korcyl5}.

        \subsection{Bosonic sectors}

        We start with the purely bosonic sector.
        In order to obtain the recurrence relation for $a_{p_2, p_3, \dots, p_N}$ we must evaluate
        the action of the Hamiltonian on a generic basis state. We get (see Appendix \ref{sec. app} for the details of calculations):
        \begin{align}
        &(a^{\dagger}a)|p_2, \dots, p_N \rangle = \frac{1}{2} \Big( \sum_{k=2}^N  k p_k \Big)|p_2, \dots, p_N \rangle, \nonumber \\
        &(a^{\dagger} a^{\dagger})|p_2,\dots, p_N \rangle = |p_2+1,p_3, \dots, p_N \rangle, \nonumber \\
        &(aa)|p_2,\dots,p_N\rangle = \Big( p_2\big( p_2 + \frac{1}{2}(N^2-1) -1 \sum_{k=3}^N k p_k\big) \Big) |p_2-1,p_3,\dots,p_N\rangle + \nonumber \\
        &+\sum_{j=3}^N \Bigg( \frac{j^2 p_j(p_j-1)}{4}\big( (a^{\dagger 2j-2})-\frac{1}{N}(a^{\dagger j-1})^2 \big)
        + \frac{j p_j}{4}\sum_{t=2}^{j-4}(a^{\dagger t})(a^{\dagger j})(a^{\dagger j-2-t})+ \nonumber \\
        &+ \frac{N j p_j}{4}\big(1-\frac{j-1}{N^2}\big)(a^{\dagger j})(a^{\dagger j-2}) + \sum_{s=j+1}^{N} \frac{j p_j s p_s}{2} \Big( \frac{(a^{\dagger j+s-2})(a^{\dagger j})}{(a^{\dagger s})} + \nonumber\\
        &- \frac{1}{N} \frac{(a^{\dagger j-1})(a^{\dagger j})(a^{\dagger s-1})}{(a^{\dagger s})} \Big) \Bigg)|p_2,\dots,p_j-2,\dots,p_N\rangle.
        \label{eq. dzialanie ada}
        \end{align}
        Eqs.\eqref{eq. dzialanie ada} lead to the general recursion relation 
        \begin{multline}
        a_{p_2-1,\dots,p_N}
         - \Big( \sum_{k=2}^N k p_k + \frac{1}{2}(N^2-1) - 2E \Big) a_{p_2,\dots,p_N}+\\
        + \Big( (p_2+1)\big(p_2 + \frac{1}{2}(N^2-1)+ \sum_{k=3}^N k p_k \big) \Big) a_{p_2+1,\dots,p_N}
        + \\
        + \sum_{j=3}^N \Bigg(
        \frac{p_j(p_j-1)}{4}j^2\Big( a_{p_2,\dots,p_j+2,\dots,p_{2j-2}-1,\dots,p_N}-\frac{1}{N} a_{p_2,\dots,p_{j-1}-2,p_j+2,\dots,p_N} \Big) +\\
        +p_j j \frac{N}{4}\big(1-\frac{j-1}{N^2}\big)a_{p_2,\dots,p_{j-2}-1,p_{j-1},p_j+1,\dots,p_N} + \\
        +\frac{p_j j}{4}\sum_{t=2}^{j-4} a_{p_2,\dots,p_t-1,\dots,p_{j-2-t}-1,\dots,p_j+1,\dots,p_N} + \\
        +p_j j \sum_{s=j+1}^{N} \frac{p_s s}{2}\Big( a_{p_2,\dots,p_j+1,\dots,p_s+1,\dots,p_{j+s-2}-1,\dots,p_N}+ \\ -
        \frac{1}{N}a_{p_2,\dots,p_{j-1}-1,p_j+1,\dots,p_{s-1}-1,p_{s}+1,\dots,p_N}\Big)
        \Bigg) = 0.
        \label{eq. dzialanie}
        \end{multline}

        An important remark concerns the Cayley-Hamilton theorem. In deriving eq.\eqref{eq. dzialanie} we did not simplified
        the operators with powers
        of bosonic creation operators bigger than $N$. Such operators can appear in two places in eq.\eqref{eq. dzialanie ada},
        namely for the operators $(a^{\dagger 2j-2})$ and $(a^{\dagger j+s-2})$, where $3\le j \le N$ and $j+1 \le N$.
        They must be reduced once the final form of the
        recurrence relation for a given $N$ is obtained. Note, however, that when examining
        the large-$N$ limit, the Cayley-Hamilton theorem does not apply, therefore,
        our recurrence relation is a good starting point for such investigation.

        With $N=3$ and after simplifying the operator $(a^{\dagger 4})$ with the Cayley-Hamilton
        theorem, we recover the recurrence relation discussed in Refs.\cite{korcyl4,korcyl5}, namely,
        \begin{multline}
        a_{p_2-1,p_3} - \big( 2p_2 + 3p_3 + 4 -2E \big) a_{p_2,p_3} + (p_2+1)(p_2+3p_3+4) a_{p_2+1,p_3} +\\+ \frac{3}{8}(p_3+1)(p_3+2) a_{p_2-2,p_3+2} = 0.
        \end{multline}
        The first three terms does not involve any change of the $p_3$ index. The mixing of amplitudes with different
        values of the $p_3$ index is described only
        by the fourth term. It is induced by the non-orthogonal states $\langle 0,2|3,0\rangle \ne 0$.

        In order to demonstrate the increasing complexity of the recursion relation with increasing $N$ induced by the
        increasing number of mixing terms between states with the same number of bosonic quanta
        but made with different elementary bricks, let us present
        the recursion relation for the $SU(4)$ model, which reads
        \begin{multline}
        a_{p_2-1,p_3,p_4} - \big( 2p_2 + 3p_3 + 4p_4 + \frac{15}{2} -2E \big) a_{p_2,p_3,p_4} +\\+ (p_2+1)(p_2+3p_3+4p_4+\frac{15}{2}) a_{p_2+1,p_3,p_4} \\
        + \big(5p_3(p_4+1)-\frac{3}{2}+3p_4(p_4+1)+\frac{13}{4}(p_4+1)\big)a_{p_2-1,p_3,p_4+1}
        +\\+ \frac{1}{3}(p_4+1)(p_4+2)a_{p_2,p_3-2,p_4+2}
        - \frac{1}{2}(p_4+1)(p_4+2)a_{p_2-3,p_3,p_4+2}
        +\\+ \frac{9}{4}(p_3+1)(p_3+2)a_{p_2,p_3+2,p_4-1}
        - \frac{9}{16}(p_3+1)(p_3+2)a_{p_2-2,p_3+2,p_4}.
        \label{eq. su4 recursion relation}
        \end{multline}
        The first line again contains terms where only the $p_2$ index vary ($p_3$ and $p_4$ remain fixed in these expressions).
        The remaining terms are mixing terms, which are induced by the nonvanishing scalar products:
        $\langle 2,0,0|0,0,1\rangle \ne 0$, $\langle 1,2,0|0,0,2\rangle \ne 0$,
        $\langle 1,0,1|0,2,0\rangle \ne 0$ and $\langle 3,0,0|0,2,0\rangle \ne 0$.

        Summarizing, a general feature of the recursion relation eq.\eqref{eq. dzialanie} is that only the first three terms
        describe a change in the $p_2$ index of the $a_{p_2,p_3\dots,p_N}$ coefficients.
        Moreover, these terms have the structure of the Laguerre recursion relation \footnote{We define
        the Laguerre polynomials $\mathcal{L}_m^{\alpha}(x)$ as the solutions of the differential equation $x y'' + (\alpha +1 -x) y' + n y = 0$
        and the orthogonality relation $\int_0^{\infty} \mathcal{L}_m^{\alpha}(x) \mathcal{L}_n^{\alpha}(x) x^{\alpha} e^{-x} dx = \delta_{m n}$.
        The polynomials $L^{\alpha}_m(x)$ are related to $\mathcal{L}^{\alpha}_m(x)$ via
        $L^{\alpha}_m(x) = \frac{\mathcal{L}^{\alpha}_m(x)}{\Gamma(m+\alpha+1)}$,
        where $\Gamma(m)$ is the Euler gamma function, $\Gamma(m+1)=m!$ for $m$ integer.}
        with $2E$ being the argument of
        the polynomials and $\sum_{k=3}^N k p_k + \frac{1}{2}(N^2-1)-1$ playing the role of their index. The additional terms
        are responsible for the mixing between states with an equal number of quanta but constructed with different elementary bricks.
        However, keeping these indices as external parameters, one can solve eq.\eqref{eq. su4 recursion relation}.

        In sections \ref{sec. spectra} and \ref{sec. solutions} we will use these observations and the general theorems
        developed in Ref.\cite{korcyl5} for
        the $SU(3)$ model, to discuss the eigenvalues and eigenstates of $H$ which solves the above recursion relations.

        \subsection{Fermionic sectors}

        Before we discuss the solutions of eq.\eqref{eq. dzialanie} let us generalize the
        above treatment for the case of fermionic sectors. Since we have
        \begin{equation}
        \big[ H, C^{\dagger}(n_B^{\alpha}, n_F, \alpha) \big] = \frac{1}{2} n_B^{\alpha} C^{\dagger}(n_B^{\alpha}, n_F, \alpha) - \frac{1}{2} \big[(aa),C^{\dagger}(n_B^{\alpha}, n_F, \alpha)\big]
        \end{equation}
        we can rewrite the eigenequation of $H$ as
        \begin{multline}
        \sum_{\alpha=1}^{d^{n_F}} \sum_{\sum_{k=2}^N k p_k \le N_{cut}-n_B^{\alpha}}  a^{\alpha}_{p_2,p_3,\dots,p_N} \
        C^{\dagger}(n_B^{\alpha}, n_F, \alpha) \ \Big( H + \frac{n_B^{\alpha}}{2} \Big)|p_2,p_3,\dots,p_N \rangle \\
        -\frac{1}{2} \sum_{\alpha=1}^{d^{n_F}} \sum_{\sum_{k=2}^N k p_k \le N_{cut}-n_B^{\alpha}} a^{\alpha}_{p_2,p_3,\dots,p_N} \big[ (aa) , C^{\dagger}(n_B^{\alpha}, n_F, \alpha) \big] |p_2,p_3,\dots,p_N \rangle = 0.
        \label{eq. dzialanie fermionowe N 1}
        \end{multline}
        The action of the Hamiltonian on the basis states has been derived in the previous section. In order to
        obtain the recurrence relation we have to compute the commutators
        $\big[ (aa) , C^{\dagger}(n_B^{\alpha}, n_F, \alpha) \big]$ for any fermionic brick and any $N$.
        We proceed in the same manner as was did for the $SU(3)$ model\cite{korcyl5}. For any fermionic brick
        $C^{\dagger}(n_B^{\alpha}, n_F, \alpha)$, the commutator with $(aa)$ will be equal to a sum of
        $n_B^{\alpha}$ terms, each of them being equal to the fermionic brick
        $C^{\dagger}(n_B^{\alpha}, n_F, \alpha)$ with one of the bosonic creation operators substituted by an annihilation operator.
        We will write them as $G^t_{\alpha}$, where the index $t$ goes from $1$ to $n_B^{\alpha}$,
        \begin{equation}
        \big[ (aa) , C^{\dagger}(n_B^{\alpha}, n_F, \alpha) \big] = \sum_{t=1}^{n_B^{\alpha}} G^t_{\alpha}.
        \end{equation}
        The operators $G^t_i$ should be now pushed on the right through the creation
        operators $\prod_{k=2}^{N} (a^{\dagger \ k})^{p_k}$. We get,
        \begin{equation}
        \forall_t \ \big[ G^t_{\alpha}, \prod_{k=2}^{N} (a^{\dagger \ k})^{p_k} \big] =
        \sum_{j=2}^N \Big(\prod_{k=2}^{j-1} (a^{\dagger k})^{p_k}\Big) \Big[ G^t_{\alpha}, (a^{\dagger j})^{p_j} \Big] \Big( \prod_{k=j+1}^{N} (a^{\dagger k})^{p_k} \Big)
        \end{equation}
        $G^t_{\alpha}$ contains exactly one bosonic annihilation operators, therefore we can write
        \begin{align}
        \forall_t \ \big[ G^t_{\alpha}, (a^{\dagger j})^{p_j} \big] &= p_j \ (a^{\dagger j})^{p_j-1} \big[ G_{\alpha}^t, (a^{\dagger j})\big].
        \end{align}
        For $j=2$, we can replace $\big[ G^t_{\alpha}, (a^{\dagger} a^{\dagger}) \big]$ in the above expression by
        $C^{\dagger}(n_B^{\alpha}, n_F, \alpha)$.
        Thus, we get the general form of the recurrence relation,
        \begin{multline}
        \sum_{\alpha=1}^{d^{n_F}} \sum_{\sum_{k=2}^N k p_k \le N_{cut}-n_B^{\alpha}}
          \Bigg\{ 
          \Bigg(
        a^{\alpha}_{p_2-1,p_3, \dots, p_N} +\\- \Big( 2p_2 + \sum_{k=3}^N k p_k + \frac{1}{2}(N^2-1) + n_B^{\alpha} - 2E \Big)a^{\alpha}_{p_2,p_3, \dots, p_N} \\
        + (p_2+1) \Big( p_2 + \sum_{k=3}^{N} k p_k  +  \frac{1}{2}(N^2-1) +n_B^{\alpha} \Big) a^{\alpha}_{p_2+1 ,p_3, \dots, p_N}  \Bigg) |\alpha,n_F;p_2,p_3,\dots,p_N \rangle \\
        + \sum_{j=3}^N  \Bigg(
        \frac{p_j(p_j-1)}{4}j^2\Big( a^{\alpha}_{p_2,\dots,p_j+2,\dots,p_{2j-2}-1,\dots,p_N}-\frac{1}{N} a^{\alpha}_{p_2,\dots,p_{j-1}-2,p_j+2,\dots,p_N} \Big) +\\
        +p_j j \frac{N}{4}\big(1-\frac{j-1}{N^2}\big)a^{\alpha}_{p_2,\dots,p_{j-2}-1,p_{j-1},p_j+1,\dots,p_N}+\\
        +\frac{p_j j}{4}\sum_{t=2}^{j-4} a^{\alpha}_{p_2,\dots,p_t-1,\dots,p_{j-2-t}-1,\dots,p_j+1,\dots,p_N} + \\
        +p_j j \sum_{s=j+1}^{N} \frac{p_s s}{2}\Big( a^{\alpha}_{p_2,\dots,p_j+1,\dots,p_s+1,\dots,p_{j+s-2}-1,\dots,p_N} + \\ -
        \frac{1}{N}a^{\alpha}_{p_2,\dots,p_{j-1}-1,p_j+1,\dots,p_{s-1}-1,p_{s}+1,\dots,p_N}\Big)
        \Bigg)  |\alpha,n_F;p_2,\dots,p_N \rangle\\
        + a^{\alpha}_{p_2,p_3, \dots, p_N} \sum_{t=1}^{n_B^{\alpha}} \Bigg(
        \sum_{k=3}^N p_k \Big[ G^t_{\alpha}, (a^{\dagger k}) \Big] |p_2,p_3,\dots,p_j-1,\dots,p_N\rangle
        +\\+ \Big( \prod_{k=2}^{N} (a^{\dagger \ k})^{p_k} \Big) G^t_{\alpha} |0 \rangle \Bigg) \Bigg\} = 0.
        \label{eq. dzialanie fermionowe N 2}
        \end{multline}
        This recursion relation can be divided into three parts. The first part (first three terms contained in a parenthesis) is diagonal
        in all indices except $p_2$, this includes also the fermionic index $\alpha$. Again, it has
        the structure of the Laguerre polynomials recursion relation with $2E$ playing the role of the argument of these polynomials.
        The second part (four terms contained in a parenthesis) mixes
        the indices $p_j$, $2\le j \le N$. However, it is still diagonal in the fermionic index $\alpha$. The third part contains terms which mix
        different values of the $\alpha$ index. It can be argued that this latter part does not contain terms proportional to the
        $C^{\dagger}(n_B^{\alpha},n_F,\alpha)$ fermionic brick, hence it really corresponds to a mixing part.

        \section{Grouping into families}

        One of the most important conclusion of this note is the observation that the solutions,
        both bosonic as well as fermionic, group into disjoint sets. This is true for the models with arbitrary $N$.
        Such sets of solutions were called \emph{families} in the study of the $SU(3)$ model\cite{korcyl5}. In this latter case,
        each solution belonged to an unique family. Families were labeled by a single integer denoting the maximal
        number of cubic bricks which appeared in the decomposition of the solutions from this family in the Fock basis. Basing on the
        form of the recursion relations eqs.\eqref{eq. dzialanie} and \eqref{eq. dzialanie fermionowe N 2}
        we now argue that these remarks can be generalized to the case of the $SU(N)$ model
        with arbitrary $N$. This is by no means a strick proof; in order to obtain a specific solution
        one should apply a general theorem presented in Ref.\cite{korcyl5}. Below, we just sketch the argument.

        We start with the discussion of the purely bosonic recursion relation.
        Let us note that at \emph{finite} cut-off there always exists one set of coefficients $a_{p_2,p_3, \dots, p_N}$
        which recursion relation is not coupled to any other. This is because
        by cutting the Fock basis we do not consider states with sufficiently large number of quanta.
        In order to be more specific let us fix the cut-off to be equal to $N_{cut}$. We choose a set of integers
        $p_3^{max}, p_4^{max}, \dots, p_N^{max}$, such that $\sum_{k=3}^N k p_k^{max} = N_{cut}$.
        One of the possible choices is $p_4^{max}=p_5^{max}=\dots=p_N^{max}=0$ and $3 p_3^{max} = N_{cut}$.
        Other choices are independent and lead to the same conclusions. Consider now the recursion
        relation for $a_{p_2,p_3^{max},p_4^{max}, \dots, p_5^{max}}$. There is only one equation which reads
        \begin{align}
        - \big( N_{cut} + \frac{1}{2}(N^2-1)-2E \big) a_{0,\frac{1}{3}N_{cut},0, \dots, 0} = 0.
        \label{eq. qc2}
        \end{align}
        Other coefficients $a_{p_2,\frac{1}{3}N_{cut},0, \dots, 0}$ are absent because they contain to many quanta.
        From eq.\eqref{eq. qc2} follow two possibilities. Either $2E = N_{cut} + \frac{1}{2}(N^2-1)$ and
        then $a_{0,\frac{1}{3}N_{cut},0, \dots, 0}$ can be arbitrary or
        $2E \ne N_{cut} + \frac{1}{2}(N^2-1)$ in which case $a_{0,\frac{1}{3}N_{cut},0, \dots, 0}$ must vanish.
        The condition for $E$ can be also rewritten in the form,
        \begin{equation}
        L_1^{N_{cut} + \frac{1}{2}(N^2-1)-1}(2E)=0 \quad \textrm{or} \quad L_1^{N_{cut} + \frac{1}{2}(N^2-1)-1}(2E) \ne 0.
        \label{eq. qc}
        \end{equation}
        Now, if the coefficient $a_{0,\frac{1}{3}N_{cut},0, \dots, 0}$ vanishes (we assume that so do
        the coefficients corresponding to other choices
        of $p_3^{max}, p_4^{max}, \dots, p_N^{max}$, such that $\sum_{k=3}^N k p_k^{max} = N_{cut}$), then the recursion relation
        for the coefficients $a_{p_2,\frac{1}{3}N_{cut}-2,0, \dots, 0}$ has no mixing terms and can be easily solved.
        The solution yields a new quantization condition of the form similar to eq.\eqref{eq. qc}, namely,
        \begin{equation}
        L_4^{N_{cut} + \frac{1}{2}(N^2-1)-7}(2E)=0.
        \label{eq. qc3}
        \end{equation}
        For $E$ satisfying the condition eq.\eqref{eq. qc3},
        nontrivial values for the coefficients $a_{p_2,\frac{1}{3}N_{cut}-2,0, \dots, 0}$ are possible.
        For $E$ which does not satisfy the condition eq.\eqref{eq. qc3}, the corresponding
        coefficients must vanish, hence
        yielding the recursion relation for another set of coefficients without mixing terms. In this way the families of solutions arise.
        For every $E$ satisfying any of the quantization conditions a new solution appear. All solutions coming from a single
        quantization condition have similar properties; in particular, the amplitudes in their decomposition in the
        Fock basis are given by Laguerre polynomials with the same index.
        Every solution can be unambiguously denoted by its energy $E$ and a set of integer numbers
        $p_3^{max}, p_4^{max}, \dots, p_N^{max}$, where $p_t^{max}$ denotes the maximal power of the elementary brick
        $(a^{\dagger t})$ in the decomposition of the eigenstate in the basis. It is a natural generalization of the
        results obtained for the $SU(3)$ model, in which case the solutions were labeled by $E$ and a single integer $p_3^{max}$.
        Hence, for example for the model with $SU(4)$ gauge symmetry, the families are labeled by two integers $p_3^{max}$ and $p_4^{max}$.

        Summarizing, for a given cut-off $N_{cut}$ the eigenenergies are given by a set of quantization conditions, all of which have the
        following form,
        \begin{equation}
        L_{\lfloor \frac{1}{2} \big( N_{cut}- (\sum_{k=3}^N k p_k) \big) \rfloor + 1}^{\sum_{k=3}^N k p_k  + \frac{1}{2}(N^2-1) -1}(2E)=0.
        \label{eq. qc4}
        \end{equation}
        Each set of numbers $\big\{ p_k \big\}$ corresponds to a quantization condition and yields a new family of solutions. The
        index of Laguerre polynomials in eq.\eqref{eq. qc4} have the structure
        \begin{equation}
        \gamma = \sum_{k=3}^N k p_k  + \frac{1}{2}(N^2-1) -1.
        \label{eq. index}
        \end{equation}
        All eigensolutions belonging to such family have decomposition coefficients in the Fock basis given by the Laguerre polynomials
        of index given by eq.\eqref{eq. index} and argument $2E$.

        By comparing the structure of the fermionic recursion relation eq.\eqref{eq. dzialanie fermionowe N 2} to the bosonic recursion
        relation eq.\eqref{eq. dzialanie} it is straightforward to generalize the above remarks to the fermionic sectors. Indeed,
        eq.\eqref{eq. dzialanie fermionowe N 2} have a part corresponding to the recursion relation in the $p_2$ index and a part corresponding
        to the mixing. One can verify that in the fermionic case the solutions of similar properties also group into families.
        They can be characterized by the index of Laguerre polynomials of the structure
        \begin{equation}
        \gamma = \sum_{k=3}^N k p_k  + \frac{1}{2}(N^2-1) -1 + n_B^{\alpha},
        \label{eq. fermionic index}
        \end{equation}
        where $n_B^{\alpha}$ describes the properties of the fermionic brick used in the decomposition of solutions in the Fock basis.
        The families are labeled by the set of integers $\big\{ p_k \big\}$ and the index $\alpha$.

        Grouping of solutions into families has two important consequences. First, as it is obvious from the above discussion,
        it allows to write explicitly
        the expressions for the spectra in all sectors for all $SU(N)$ models. This follows from the
        fact that each family has its own quantization condition, all of them having a similar structure.
        We discuss this in more details in section \ref{sec. spectra}.
        Second, one is able to consider solutions belonging to a single family, and therefore in the simplest cases
        write the solutions explicitly for any $N$. This feature is discussed in section \ref{sec. solutions}.

        \section{Spectra}
        \label{sec. spectra}

        In this section we present closed formulae describing the spectra of the SYMQM models with arbitrary $SU(N)$ gauge symmetry.

        \subsection{Bosonic sectors}

        For a given cut-off $N_{cut}$, we can define a polynomial $\Theta_{N_{cut}}^{n_F=0}(N,E)$,
        whose zeros correspond to all eigenenergies of the cut Hamiltonian operator in the bosonic sector,
        \begin{equation}
        \Big\{ E \Big\}^{n_F=0}_{N_{cut}} = \Big\{ \Theta_{N_{cut}}^{n_F=0}(N,E) = 0\Big\}.
        \end{equation}
        $\Theta_{N_{cut}}^{n_F=0}(N,E)$ can be expressed as a product of quantization conditions of all
        nonempty families of solutions. For example, for the $SU(4)$ model we have
        \begin{equation}
        \Theta_{N_{cut}}^{n_F=0}(4,E) = \prod_{t=0}^{\lfloor \frac{1}{3} N_{cut} \rfloor} \Big(
        \prod_{k=0}^{\lfloor \frac{1}{4} ( N_{cut}- 3t ) \rfloor}
        L_{\lfloor \frac{1}{2} ( N_{cut}-3t-4k ) \rfloor + 1}^{3t+4k + \frac{1}{2}(16-1) -1}(2E) \Big),
        \end{equation}
        whereas for the $SU(5)$ model
        \begin{multline}
        \Theta_{N_{cut}}^{n_F=0}(5,E) =\\= \prod_{t=0}^{\lfloor \frac{1}{3} N_{cut} \rfloor} \Big(
        \prod_{k=0}^{\lfloor \frac{1}{4} ( N_{cut}- 3t ) \rfloor} \Big(
        \prod_{s=0}^{\lfloor \frac{1}{5} ( N_{cut}- 3t -4k ) \rfloor}
        L_{\lfloor \frac{1}{2} ( N_{cut}-3t-4k-5s ) \rfloor + 1}^{3t+4k+5s + \frac{1}{2}(25-1) -1}(2E) \Big) \Big).
        \end{multline}

        These formulas were checked with independent numerical calculations which exploited a recursive algorithm\cite{korcyl1}.
        With both methods results for $N_{cut} \le 20$ were obtained and agreed exactly.

        From the recursion relation eq.\eqref{eq. dzialanie} follows a general formula for the
        polynomial $\Theta_{N_{cut}}^{n_F=0}(N,E)$ for any $N$,
        \begin{equation}
        \Theta_{N_{cut}}^{n_F=0}(N,E) =
        \prod_{i=3}^N  \Bigg( \prod_{p_i=0}^{\lfloor \frac{1}{i} \big( N_{cut}- (\sum_{k=3}^{i-1} k p_k) \big) \rfloor}
        L_{\lfloor \frac{1}{2} \big( N_{cut}- (\sum_{k=3}^N k t_k) \big) \rfloor + 1}^{\sum_{k=3}^N k p_k  + \frac{1}{2}(N^2-1) -1}(2E) \Bigg).
        \label{eq. widmo bozonowe sun}
        \end{equation}

        Once the whole spectrum can be calculated thanks to eq.\eqref{eq. widmo bozonowe sun}
        one should be able to compute the Witten index or the microcanonical partition function
        of the SYMQM systems.

        From the properties of the Laguerre polynomials, one can conclude that the smallest eigenvalues
        will belong to the family with the smallest index and the biggest order. These conditions can be satisfied by
        setting $\sum_k k p_k = 0$. Hence, the smallest eigenenergies will belong to the simplest family $\big\{p_k\big\}=\big\{0,0,\dots,0\big\}$,
        the one without any mixing.


        \subsection{Fermionic sectors}

        The generalization of the above results to the fermionic sectors is immediate. The polynomial $\Theta_{N_{cut}}^{n_F}(N,E)$ valid in all
        sector for any $SU(N)$ model can be obtained, basing on the recursion relation eq.\eqref{eq. dzialanie fermionowe N 2}, in the form
        \begin{multline}
        \Theta_{N_{cut}}^{n_F}(N,E) = \prod_{\alpha=1}^{d^{n_F}(N)} \Bigg\{
        \prod_{i=3}^N  \Bigg( \\ \prod_{p_i=0}^{\big\lfloor \frac{1}{i} \big( N_{cut}- (\sum_{k=3}^{i-1} k p_k) - n_B^{\alpha}(N) \big) \big\rfloor}
        L_{\big\lfloor \frac{1}{2} \big( N_{cut}- (\sum_{k=3}^N k p_k) - n_B^{\alpha}(N) \big) \big\rfloor + 1}^{(\sum_{k=3}^N k p_k)  + \frac{1}{2}(N^2-1) -1 + n_B^{\alpha}(N)}(2E) \Bigg) \Bigg\},
        \label{eq. widmo fermionowe sun}
        \end{multline}
        where the numbers $d^{n_F}(N)$ and $n_B^{\alpha}(N)$ depend on $N$.

        \subsection{Continuum limit}

        The continuum limit can be simply obtained by taking the limit $N_{cut} \rightarrow \infty$
        in the expressions eq.\eqref{eq. widmo bozonowe sun} and eq.\eqref{eq. widmo fermionowe sun}. The
        eigenvalues form a dense subset of the positive real numbers. Each eigenenergy is infinitely
        degenerate. This does not concern the non-degenerate supersymmetric vacua which
        will be shown when discussing the wave-functions of the supersymmetric vacua in the next section.
        The continuum limit involving the scaling law in the spirit of Refs.\cite{maciek1,korcyl0} will be
        discussed elsewhere.

        \section{Discussion of the simplest solutions}
        \label{sec. solutions}

%

        In Ref.\cite{maciek_lie_groups} Trzetrzelewski has formulated an algorithm for finding the sectors of the SYMQM systems where
        the supersymmetric vacua are located. Using the recursion relation eq.\eqref{eq. dzialanie} we can construct the wavefunctions of these vacua by taking the limit of
        $E \rightarrow 0$ of our solutions. However, in order to be able to take this limit, first one has to consider the solutions
        with $N_{cut}\rightarrow \infty$. We discuss these two limits below.

        \subsection{Solutions at finite cut-off}

        Let us start by describing the solutions at finite cut-off.
        The most general solution has the following form
        \begin{multline}
        |E,p_3,p_4,\dots,p_N\rangle^{n_F=0} =\\= e^{-E}\sum_{n=0}^{d-1}L_n^{\frac{1}{2}(N^2-1)+\sum_{s=3}^N k p_k -1}(2E)
        \Big(|n,p_3,p_4,\dots,p_N\rangle + \\
        +\sum_{t_3,t_4,\dots,t_N=1}^{p_3,p_4,\dots,p_N} A^{p_3,p_4,\dots,p_N}_{t_3,t_4,\dots,t_N} |n+\sum_{s=3}^N s t_s,p_3-t_3,p_4-t_4,\dots,p_N-t_N\rangle \Big).
        \label{eq. rozwiazanie}
        \end{multline}
        The parameter $d$ in the above sum is equal to the number of solutions belonging to the family denoted by
        $\big\{ p_3,p_4,\dots,p_N\big\}$ at finite cut-off. Obviously $d$ must depend on the cut-off, $d=d(N_{cut})$. The coefficients
        $A^{p_3,p_4,\dots,p_N}_{t_3,t_4,\dots,t_N}$ must be determined from the recursion relation eq.\eqref{eq. dzialanie}. These
        amplitudes depend only on the set of integers $\big\{p_k\big\}$; especially, they do not depend on $N_{cut}$.

        The simplest solutions belong to the family $\big\{0,0,\dots,0\big\}$. In this case $A^{p_3,p_4,\dots,p_N}_{t_3,t_4,\dots,t_N}\equiv 0$,
        i.e. the solutions have no mixing and hence are only built out of bilinear bosonic bricks. They can
        be written as
        \begin{equation}
        |E_m,0,\dots,0\rangle^{n_F=0} = e^{-E_m} \sum_{n=0}^{d_0-1} L_n^{\frac{1}{2}(N^2-1)-1}(2E_m) |n,0,\dots, 0\rangle, \ 1 \le m \le d_0,
        \label{eq. rozwiazanie f0}
        \end{equation}
        where $d_0 = \lfloor \frac{N_{cut}}{2} \rfloor + 1$, and $E_m$ are such that $L_{d_0}^{\frac{1}{2}(N^2-1)-1}(2E_m) = 0$.
        As an example of solutions belonging to more complex families, we present solutions
        from the family $\big\{2,0,\dots,0\big\}$ valid for any $N>3$. They read
        \begin{multline}
        |E_m,2,0,\dots,0\rangle^{n_F=0} = e^{-E_m}\sum_{n=0}^{d_1-1}L_n^{\frac{1}{2}(N^2-1)+5}(2E_m)
        \Big(|n,2,0,\dots,0\rangle +\\- \frac{18}{N}\frac{1}{24+6(N^2-1)}|n+3,0,0,\dots,0\rangle \Big), \qquad 1 \le m \le d_0.
        \label{eq. rozwiazanie 1}
        \end{multline}
        $d_1$ denotes the number of solutions of this type for a given cut-off and is given by $d_1 = \big\lfloor \frac{1}{2} (N_{cut} - 6) \big\rfloor$
        and $E_m$ are such that $L_{d_1}^{\frac{1}{2}(N^2-1)+5}(2E_m) = 0$ this time.

        States from other sectors can be easily obtained by acting with the fermionic
        bricks on eqs.\eqref{eq. rozwiazanie f0} or \eqref{eq. rozwiazanie 1}.
        If one uses purely fermionic bricks, then no mixing between fermionic bricks appears. Therefore, the states presented below are solutions of
        the recursion relation eq.\eqref{eq. dzialanie fermionowe N 2}, 
        \begin{multline}
        |E_m,0,\dots,0\rangle^{n_F} =\\= e^{-E_m} \sum_{n=0}^{d_0-1} L_n^{\frac{1}{2}(N^2-1)-1}(2E_m)
        (f^{\dagger 3})^{i_3}(f^{\dagger 5})^{i_5}\dots(f^{\dagger (2N-1)})^{i_{2N-1}}|n,0,\dots, 0\rangle,
        \label{eq. rozwiazanie f3}
        \end{multline}
        where $i_3,i_5,\dots,i_{2N-1} \in \{0,1\}$ and $n_F=\sum_{k=2}^{N} (2k-1) i_{2k-1}$.
        Such states can be find in the spectrum of all $SU(N)$ models (see also Ref.\cite{maciek_lie_groups}).
        Similarly,
        \begin{multline}
        |E_m,2,0,\dots,0\rangle^{n_F} =\\= e^{-E_m}\sum_{n=0}^{d_1-1}L_n^{\frac{1}{2}(N^2-1)+5}(2E_m)(f^{\dagger 3})^{i_3}(f^{\dagger 5})^{i_5}\dots(f^{\dagger (2N-1)})^{i_{2N-1}} \times \\
        \times \Big(|n,2,0,\dots,0\rangle-\frac{18}{N}\frac{1}{24+6(N^2-1)}|n+3,0,0,\dots,0\rangle \Big),
        \label{eq. rozwiazanie 2}
        \end{multline}
        with $n_F=\sum_{k=2}^{N} (2k-1) i_{2k-1}$.

        In this way expressions for increasingly complicated solutions can be obtained. 
        \subsection{Continuum limit}

        Similarly to the case of the $SU(3)$ model, the continuum limit can be easily obtained from the finite-cut-off solutions.
        The only quantity dependent on $N_{cut}$ in eqs.\eqref{eq. rozwiazanie f0} and \eqref{eq. rozwiazanie f3}
        and in eqs.\eqref{eq. rozwiazanie 1} and \eqref{eq. rozwiazanie 2} are the upper limits of the sums, $d_0$ and $d_1$, and the
        eigenenergies $E_m$.
        Hence, the continuum limit of these solutions can be simply obtained by extending the sums to infinity,
        \begin{align}
        |E,0,\dots,0\rangle^{n_F=0} &= e^{-E} \sum_{n=0}^{\infty} L_n^{\frac{1}{2}(N^2-1)-1}(2E) |n,0,\dots, 0\rangle, \nonumber \\
        |E,2,0,\dots,0\rangle^{n_F=0} &= e^{-E}\sum_{n=0}^{\infty}L_n^{\frac{1}{2}(N^2-1)+5}(2E)
        \Big(|n,2,0,\dots,0\rangle +\nonumber \\&- \frac{18}{N}\frac{1}{24+6(N^2-1)}|n+3,0,0,\dots,0\rangle \Big),\nonumber
        \end{align}
        and adequately for the fermionic solutions. In the continuum limit the set of eigenenergies, $\big\{E\big\}$, is dense in the set of real,
        positive numbers.

        \subsection{Normalization}

        The normalization of the presented states can be calculated for the simplest solutions eqs.\eqref{eq. rozwiazanie f0}
        and \eqref{eq. rozwiazanie f3}. For the continuum states we have
        \begin{multline}
        \langle E|E' \rangle = \frac{\Delta e^{-E-E'}}{\Gamma\big(\frac{1}{2}(N^2-1)\big)} \times \\ \times \lim_{z \rightarrow 1^-}
        \frac{1}{1-z} e^{\frac{-2z}{1-z}(E+E')}\big( 4 E E' z\big)^{-\frac{1}{2}(\frac{1}{2}(N^2-1)-1)} I_{\frac{1}{2}(N^2-1)-1}\Big(\frac{4\sqrt{E E' z}}{1-z} \Big),
        \end{multline}
        where
        \begin{equation}
        \Delta = \left\{ \begin{array}{ll}
        &\langle0|0\rangle=1 \textrm{ in the bosonic sector} \nonumber \\
        &\langle0|(f^{2N-1})^{i_{2N-1}}\dots(f^5)^{i_5}(f^{3})^{i_3}(f^{\dagger 3})^{i_3}(f^{\dagger 5})^{i_5}\dots(f^{\dagger (2N-1)})^{i_{2N-1}}|0\rangle \\
        &\qquad \qquad \qquad \qquad \textrm{ in the sector with } n_F=\sum_{k=2}^{N} (2k-1) i_{2k-1}.
        \end{array} \right.
        \end{equation}
        Expressing $z$ as $z=1-4\epsilon$ we get
        \begin{equation}
        \langle E|E' \rangle = \frac{\Delta}{2\sqrt{\pi}}\frac{e^{E+E'-2\sqrt{E E'}}}{\Gamma\big(\frac{1}{2}(N^2-1)\big)} \big( 4 E E'\big)^{-\frac{1}{2}(\frac{1}{2}(N^2-1)-1)-\frac{1}{4}}
        \lim_{\epsilon \rightarrow 0^+}
        \frac{1}{\sqrt{4\epsilon}}e^{-\frac{\big(\sqrt{2E} - \sqrt{2E'}\big)^2}{4 \epsilon}}.
        \end{equation}
        Exploiting the well-known representation of Dirac delta distribution we have
        \begin{align}
        \langle E|E' \rangle &= \frac{\Delta}{2\sqrt{\pi}}\frac{e^{E+E'-2\sqrt{E E'}}}{\Gamma\big(\frac{1}{2}(N^2-1)\big)}
        \big( 4 E E'\big)^{-\frac{1}{4}(N^2-1)+\frac{1}{4}}
        \frac{\sqrt{\pi}}{2} \delta\big( \sqrt{2E} - \sqrt{2E'} \big) \nonumber \\
        &= \frac{\Delta}{4\Gamma\big(\frac{1}{2}(N^2-1)\big)}
        \big( 2 E\big)^{-\frac{1}{2}(N^2-2)} \delta\big( \sqrt{2E} - \sqrt{2E'} \big).
        \label{eq. normalizacja}
        \end{align}
        Hence, the continuum solutions belonging to these simplest solutions are orthogonal to each other and normalized as plane-waves.
        It may by explicitly checked that they are also orthogonal to solutions of any other family.

        The presence of the factor $\big( 2 E\big)^{-\frac{1}{2}(N^2-2)}$ may be linked with the jacobian of the change of variables from
        the 'cartesian' variables to the 'spherical' variables. One can think of a set of 'Fourier transformed' degrees of freedom denoted
        by $k_A$ corresponding to the original set of degrees of freedom $\phi_A$. The normalization factor then depends only
        on the 'radial' variable, $(2E)^2 = \sum_A k_A^2$. The explicit form of the 'Fourier' transformation
        appropriate for the $SU(N)$ manifolds is not known. It is therefore surprising that, the $SU(N)$ manifold
        being $N^2-1$ dimensional, the mentioned factor in eq.\eqref{eq. normalizacja} corresponds exactly
        to the jacobian of the change of variables in a $N^2-1$ dimensional Euclidean space.

        \subsection{Vacuum solutions}

        Imposing the normalization of all solutions according to eq.\eqref{eq. normalizacja} has an important consequence.
        It turns out that solutions
        belonging to more complicated families $\big\{p_3, p_4, \dots, p_N;\alpha\big\}$ with at least one of $p_i > 0$ acquire an normalization factor
        of the form $(2E)^{\gamma}$, where $\gamma$ is a positive real number. This implies, that in the limit of $E \rightarrow 0$, these
        solutions vanish. Hence, the supersymmetric vacua can exist only in the sectors where the
        simplest families $\big\{0,0,\dots,0; \alpha\big\}$ can be constructed. These are exactly the same sectors as those found
        in Ref.\cite{maciek_lie_groups} where these conclusions were reached from the cohomology of the
        $SU(N)$ groups point of view.

        Finally, let us note that the fundamental theorem of supersymmetry, namely
        \begin{equation}
        Q|\textrm{vacuum}\rangle = Q^{\dagger}|\textrm{vacuum}\rangle = 0 \Leftrightarrow \langle \textrm{vacuum} | H | \textrm{vacuum} \rangle = 0,
        \label{eq. susy theorem}
        \end{equation}
        may not hold when the vacuum state, $|\textrm{vacuum}\rangle$, is not normalizable.
        Indeed, in the cases discussed above, $Q|\textrm{vacuum}\rangle \ne 0$ since the state $Q|\textrm{vacuum}\rangle$ exists and
        can be calculated. However, with the normalization eq.\eqref{eq. normalizacja}
        the state $Q|\textrm{vacuum}\rangle$ has zero norm. In this situation, eq.\eqref{eq. susy theorem} is valid and
        all the vacua are the true, non-degenerate, supersymmetric ground states.

        \subsection{Completeness}


        We emphasized the fact that solutions can be divided into families. The transformation from the Fock states to the solutions
        of a given family was shown to be non-degenerate\cite{korcyl5}. Hence, one can show that the transformation of the entire Fock basis
        onto the set of all solutions is also non-degenerate. Therefore, since the Fock basis was shown to span the entire Hilbert
        space of the SYMQM models, equivalently the set of all solutions is also complete.

        \section{Conclusions}

        In this work we generalized the solutions, which were recently derived for the $SU(3)$ model, to the
        general case of models with $SU(N)$ gauge groups.
        The new solutions have all the required properties: the are orthogonal, complete and normalized according to the plane-wave normalization.
        Hence, we obtained the correct $SU(N)$ generalization of the Claudson-Halpern solutions derived for the $SU(2)$ model.

        After deriving the recursion relations for the amplitudes in the decomposition of the eigenstates in the Fock basis we discussed
        the properties of their solutions. We have emphasized that these solutions group into disjoint sets called families. Within each
        family the possible eigenenergies can be calculated from a single quantization condition which involves a Laguerre polynomial
        with a particular index $\alpha$. Moreover, we argued that the amplitudes of all solutions belonging to a given family are given
        by Laguerre polynomials with the same index $\alpha$.

        In section \ref{sec. spectra} we provided closed formulae for the spectra of the studied models. These expressions
        are valid for any cut-off $N_{cut}$ and for any gauge group $SU(N)$. They may be subsequently used in the studies
        of the thermodynamics of the SYMQM models (for a recent article on the thermodynamics of higher dimensional SYMQM models
        see Ref\cite{nishimura}).

        In section \ref{sec. solutions} we discussed the properties of
        the simplest solutions. We explicitly proved the orthogonality of solutions
        belonging to the $\big\{0,\dots,0\big\}$ family as well as determined the normalization
        factors of these solutions. Consequently we could confirm the construction of
        supersymmetric vacua presented in Ref.\cite{maciek_lie_groups} and we provided
        their correct wave-functions. However, due to the described normalization of
        our solutions, we did not had to impose any compactification as was done in Ref.\cite{maciek_lie_groups}.

        Our results can be immediately used in the studies of the large-N limit of the SYMQM models. The large-N limit of systems
        in a Fock representation was studied by Thorn\cite{thorn} and was recently reexamined in the context of a
        supersymmetric quantum mechanical model
        in Refs.\cite{vw1,vw2,vw3}. Additionally, the large-N limit of wave-functions of higher dimensional SYMQM are particularly interesting
        since they can have an interpretation in the context of supermembrane theory \cite{maciek_spiky}.
        The expressions derived in section \ref{sec. solutions} with their explicit $N$ dependence may provide some new insight on this limit.

        \section*{Acknowledgements}

The Author would like to acknowledge many useful discussions with prof. J. Wosiek on the subject of this paper.

        \appendix

        \section{Calculations leading to eq.\eqref{eq. dzialanie ada}}
        \label{sec. app}

        In order to evaluate the action of the $(aa)$ operator one has to move it through all operators composing the basis
        state to the point when it hits the Fock vacuum. Hence
        \begin{align}
        (aa)|p_2,p_3,p_4, \dots, p_N \rangle = \sum_{j=2}^N \Big(\prod_{i=2}^{j-1} (a^{\dagger i})^{p_i}\Big) \Big[ (aa), (a^{\dagger j})^{p_j} \Big] \Big( \prod_{i=j+1}^{N} (a^{\dagger i})^{p_i} \Big) |0 \rangle
        \end{align}
        We need the following commutators
        \begin{equation}
        \Big[(aa), (a^{\dagger n}) \Big] = n (a^{\dagger n-1 a}) + \frac{n N}{2}\big( \frac{1}{2} - \frac{n-1}{2N^2}\big)(a^{\dagger n-2})
        + \frac{n}{4} \sum_{j=2}^{n-4} (a^{\dagger j})(a^{\dagger n-2-j}).
        \end{equation}
        Hence,
        \begin{multline}
        \Big[(aa), (a^{\dagger n})^{m} \Big] = \frac{1}{4}m(m-1)n^2(a^{\dagger n})^{m-2}(a^{\dagger 2n-2})+\\
        -\frac{1}{4}m(m-1)\frac{n^2}{N}(a^{\dagger n})^{m-2}(a^{\dagger n-1})^2
        + m n (a^{\dagger n})^{m-1}(a^{\dagger n-1}a)+\\
        + m n \frac{N}{2} \big( \frac{1}{2} - \frac{n-1}{2N^2} \big) (a^{\dagger n-2})(a^{\dagger n})^{m-1} 
        + \frac{m n}{4} \sum_{j=2}^{n-4} (a^{\dagger j})(a^{\dagger n-2-j})(a^{\dagger n})^{m-1}
        \end{multline}
        and for any positive integers $A$ and $B$
        \begin{equation}
        \Big[(a^{\dagger n}a), (a^{\dagger m})^{k} \Big] = k \Big( \frac{m}{2}(a^{\dagger n+m-1})-\frac{m}{2N}(a^{\dagger n})(a^{\dagger m-1})\Big) (a^{\dagger m})^{k-1}
        \end{equation}
        \begin{multline}
        \Big[ (a^{\dagger n-1}a), \Big( \prod_{i=n+1}^N (a^{\dagger i})^{p_i}\Big) \Big] =\\=
        \Big(\prod_{i=n+1}^N (a^{\dagger i})^{p_i} \Big) \sum_{t=n}^{N-1} \frac{p_{i+1}(i+1)}{2} \Bigg( \frac{(a^{\dagger n+i-1})}{(a^{\dagger i+1})}
        - \frac{1}{N} \frac{(a^{\dagger n-1})(a^{\dagger i})}{(a^{\dagger i+1})} \Bigg)
        \end{multline}
        where we have introduce the notation
        \begin{equation}
        \prod_{i=A}^B \frac{(a^{\dagger i})^{p_i}}{(a^{\dagger t})^n} \equiv \Big( \prod_{i=A}^{t-1} (a^{\dagger i})^{p_i} \Big) (a^{\dagger t})^{p_t-n} \Big( \prod_{i=t+1}^{B} (a^{\dagger i})^{p_i} \Big).
        \end{equation}
        Therefore,
        \begin{multline}
        \Big[ (aa), \Big( \prod_{j=2}^N (a^{\dagger j})^{p_j}\Big) \Big]|0\rangle =\\= \Big[ p_2\big( p_2 + \frac{1}{2}(N^2-1) -1 \sum_{s=3}^N s p_s\big) \Big] \frac{\Big( \prod_{j=2}^N (a^{\dagger j})^{p_j}\Big)}{(a^{\dagger 2})} |0\rangle +  \\
        +\sum_{j=3}^N \Bigg[ \frac{j^2 p_j(p_j-1)}{4}\big( (a^{\dagger 2j-2})-\frac{1}{N}(a^{\dagger j-1})^2 \big)
        + \frac{j p_j}{4}\sum_{t=2}^{j-4}(a^{\dagger t})(a^{\dagger j})(a^{\dagger j-2-t})+  \\
        + \frac{N j p_j}{4}\big(1-\frac{j-1}{N^2}\big)(a^{\dagger j})(a^{\dagger j-2}) + \sum_{s=j+1}^{N} \frac{j p_j s p_s}{2} \Big( \frac{(a^{\dagger j+s-2})(a^{\dagger j})}{(a^{\dagger s})} + \\
        - \frac{1}{N} \frac{(a^{\dagger j-1})(a^{\dagger j})(a^{\dagger s-1})}{(a^{\dagger s})} \Big) \Bigg] \frac{\Big( \prod_{j=2}^N (a^{\dagger j})^{p_j}\Big)}{(a^{\dagger j})^2}  |0\rangle.
        \end{multline}


\end{document}